\documentclass[%
reprint,
superscriptaddress,
showpacs,preprintnumbers,
amsmath,amssymb,
aps,
pra,
longbibliography
]{revtex4-2}
\usepackage{color}
\usepackage{graphicx}
\usepackage{bm}
\usepackage{mathrsfs}
\usepackage{dsfont}
\usepackage{ulem}
\usepackage[colorlinks,allcolors=blue]{hyperref}

\makeatletter

\newcommand{\Rmnum}[1]{\expandafter\@slowromancap\romannumeral #1@}
\newcommand{\mi}{i}
\makeatother

\begin{document}
\title{Realization of a deterministic quantum Toffoli gate with a single photon}

\author{Shihao Ru}
\affiliation{Shaanxi Key Laboratory of Quantum Information and Quantum Optoelectronic Devices, School of Physics of Xi'an Jiaotong University, Xi'an 710049, China}

\author{Yunlong Wang}
\email{yunlong.wang@mail.xjtu.edu.cn}
\affiliation{Shaanxi Key Laboratory of Quantum Information and Quantum Optoelectronic Devices, School of Physics of Xi'an Jiaotong University, Xi'an 710049, China}

\author{Min An}
\affiliation{Shaanxi Key Laboratory of Quantum Information and Quantum Optoelectronic Devices, School of Physics of Xi'an Jiaotong University, Xi'an 710049, China}

\author{Feiran Wang}
\affiliation{School of Science of Xi’an Polytechnic University, Xi'an 710048, China}

\author{Pei Zhang}
\affiliation{Shaanxi Key Laboratory of Quantum Information and Quantum Optoelectronic Devices, School of Physics of Xi'an Jiaotong University, Xi'an 710049, China}

\author{Fuli Li}
%\email{flli@xjtu.edu.cn}
\affiliation{Shaanxi Key Laboratory of Quantum Information and Quantum Optoelectronic Devices, School of Physics of Xi'an Jiaotong University, Xi'an 710049, China}
\date{\today}

\begin{abstract}
Quantum controlled-logic gates, including controlled {\scriptsize NOT} gate and Toffoli gate, play critical roles in lots of quantum information processing schemes.
We design and experimentally demonstrate deterministic Toffoli gate by utilizing orbital-angular-momentum and polarization degrees of freedom of a single photon. In addition, we generate Bell states by using the controlled {\scriptsize NOT} gate. The effective conversion rate of the Toffoli gate in our experiment is $(95.1\pm3.2)\%$.
Furthermore, our experimental setup does not require any auxiliary photons and probabilistic post selections.
\end{abstract}

\maketitle
\section{Introduction} % (fold)
\label{sec:introduction}
Realization of universal quantum logic gates, the core requirement of quantum computers, has attracted widespread attention \cite{2017PRLhdgates,sekiguchi2017optical,2018PRLsunSC,nagata2018universal,2020PRLhdgates}.
There are two different universal discrete sets of quantum logic gates \cite{nielsen_chuang_2010,boykin1999universal,quantumgate1995,2005Cliffordgate}.
The basic set consists of Hadamard, phase-flip, and $\pi/8$ gates of single qubit, and controlled {\scriptsize NOT} ({\scriptsize CNOT}) gate of two qubits.
The other set consists of the Hadamard, phase-flip, controlled-{\scriptsize NOT} and Toffoli gates \cite{2013Toffoli}.

Multi-qubit gates, of which the Toffoli gate is one, are essential elements in quantum networks, quantum simulations, and quantum algorithms, and they serve as stepping stones to implement scalable quantum computers.
Usually with some physical platforms, including linear optics \cite{2006LinearopticsToffoli,lanyon2009simplifying}, trapped ions \cite{2009TrappedionsToffoli}, superconducting circuits \cite{fedorov2012implementation}, and neutral atoms \cite{2019NeutralAtoms}, one has proposed and experimentally demonstrated Toffoli-gate schemes with a single degree of freedom (DoF).
The physical realization of Toffoli gates, whether requiring auxiliary qubits and probabilistic post selection \cite{2006LinearopticsToffoli,2013tofolli_lab} or simplifying the Toffoli gate with five two-qubit gates \cite{2013Toffoli} (the decomposition of a generalized $n$-qubit Toffoli gate requires nearly $O(n^2)$ two-qubit gates), needs more quantum resources.
The inefficient synthesis of Toffoli gates increases the length and time scales of quantum circuits, and makes the gates further susceptible to their environment  \cite{2013Toffoli}.

Therefore, utilizing as few quantum resources as possible to realize quantum logic gates and performing quantum computing is important.
Two strategies, in order to solve this problem, are commonly adopted.
One is to exploit a system with other accessible high-dimensional states (qudits) in quantum information processing \cite{erhard2018twisted,forbes2019quantum,erhard2020advances}.
Another is to encode qubits in multiple DoFs of a quantum system \cite{huang2017deterministic,starek2018experimental,starek2016experimental}.
Compared to multi-photon single-DOF qubits, some positive operator-valued measurements can be easily achieved in the multi-DoF qubits \cite{wang2015quantum,li2019experimental}.
As usual, extra DoF-qubits are used as auxiliary qubits in quantum information processing.
Both enhance channel capacity, allow less-decoherence quantum information processing with as fewer quantum resources as possible, and the fault tolerance and noise resistance of high-dimensional states are much better \cite{securitydlevel2002,lo2014secure,PRX2019noise}.
Additionally, in the above strategies, the transverse spatial mode DoF, due to its high-dimensional quantum properties, various effective control, and identification technologies \cite{rubinsztein2016roadmap,erhard2018twisted,shen2019optical}, has attracted a lot of attention.
One of popular and effective discretization of transverse spatial mode DoF is the Laguerre-Gauss (LG) modes (a complete and orthonormal basis) \cite{rubinsztein2016roadmap,erhard2018twisted}.
LG modes are characterized by a twisted helical wavefront with the form $\exp(\mathrm{i}\ell\phi)$, where $\phi$ is the azimuthal coordinate, and $\ell$ corresponds to the quantized orbital angular momentum (OAM) value \cite{Allen1992}.

In this research, we focus on the experimental realization of deterministic quantum Toffoli gate with a single photon.
In Sec.~\ref{sec:Experimental_setup}, we design and demonstrate a simple and feasible scheme of three-qubit Toffoli gate by exploiting the spin angular momentum (SAM) and OAM DoFs of a single photon, which can be extended to a scheme of $n$-qubit Toffoli gate.
We use a series of specific input states to experimentally illustrate the performance of our Toffoli gate in Sec.~\ref{sec:measure}.
Furthermore, we show the {\scriptsize CNOT} gate in the four-dimensional OAM-encoded space, which is the main component of the Toffoli gate, and perform quantum tomography of all output states with an average fidelity of more than $90\%$.
Our demonstrations open up new paths for quantum computing using hybrid-DoF and high-dimensional systems.
% section introduction (end)

\section{Experimental setup} % (fold)
\begin{figure}[!t]
\centering
\includegraphics[width=0.75\linewidth]{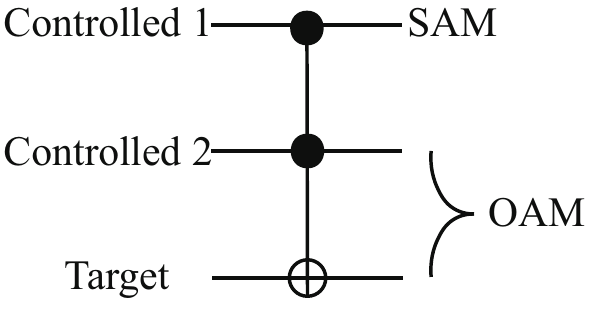}
\caption{Quantum Toffoli gate for three qubits encoded into SAM and OAM DoFs of a single photon.}\label{circuit}
\end{figure}

\label{sec:Experimental_setup}
The operation of three-qubit Toffoli gate can be written as $U_{\mathrm{Toff}}\left|x,y,z\right>=\left|x,y,z\oplus xy\right>$, where $x$, $y$, and $z$ are the binary variables and $\oplus$ denotes addition modulo two. A $n$-qubit Toffoli gate is defined to have $n-1$ controlled qubits and performs a conditional {\scriptsize NOT} gate on the $n$-th qubit (the target) if and only if all the controls are 1, which has a similar form as three-qubit Toffoli gate.
In our research, the coding scheme is illustrated in Fig.~\ref{circuit}.
The first controlled qubit is encoded by the SAM
\begin{align}
\left|{V}\right>\to\left|{0}\right>,\qquad \left|{H}\right>\to\left|{1}\right>,
\end{align}
and the second controlled qubit and the target qubit are encoded as
\begin{align}
&\left|{\ell=-2}\right>\to\left|{00}\right>,\qquad \left|{\ell=0}\right>\to\left|{01}\right>,\notag\\
&\left|{\ell=-1}\right>\to\left|{10}\right>,\qquad \left|{\ell=+1}\right>\to\left|{11}\right>
\end{align}
in a four-dimensional Hilbert space spanned by OAM quantum states.
It should be noted that our proposal can extend to the $n$-qubit Toffoli gate.
The first controlled qubit is encoded by SAM, $2^{n-2}$ even OAMs are used to encode $\left|00\cdots00\right>\to\left|11\cdots01\right>$, and a couple of odd OAMs ($\ell=\pm1$) are used to encode $\left|11\cdots10\right>$ and $\left|11\cdots11\right>$.
\begin{figure}[!b]
\centering
\includegraphics[width=\linewidth]{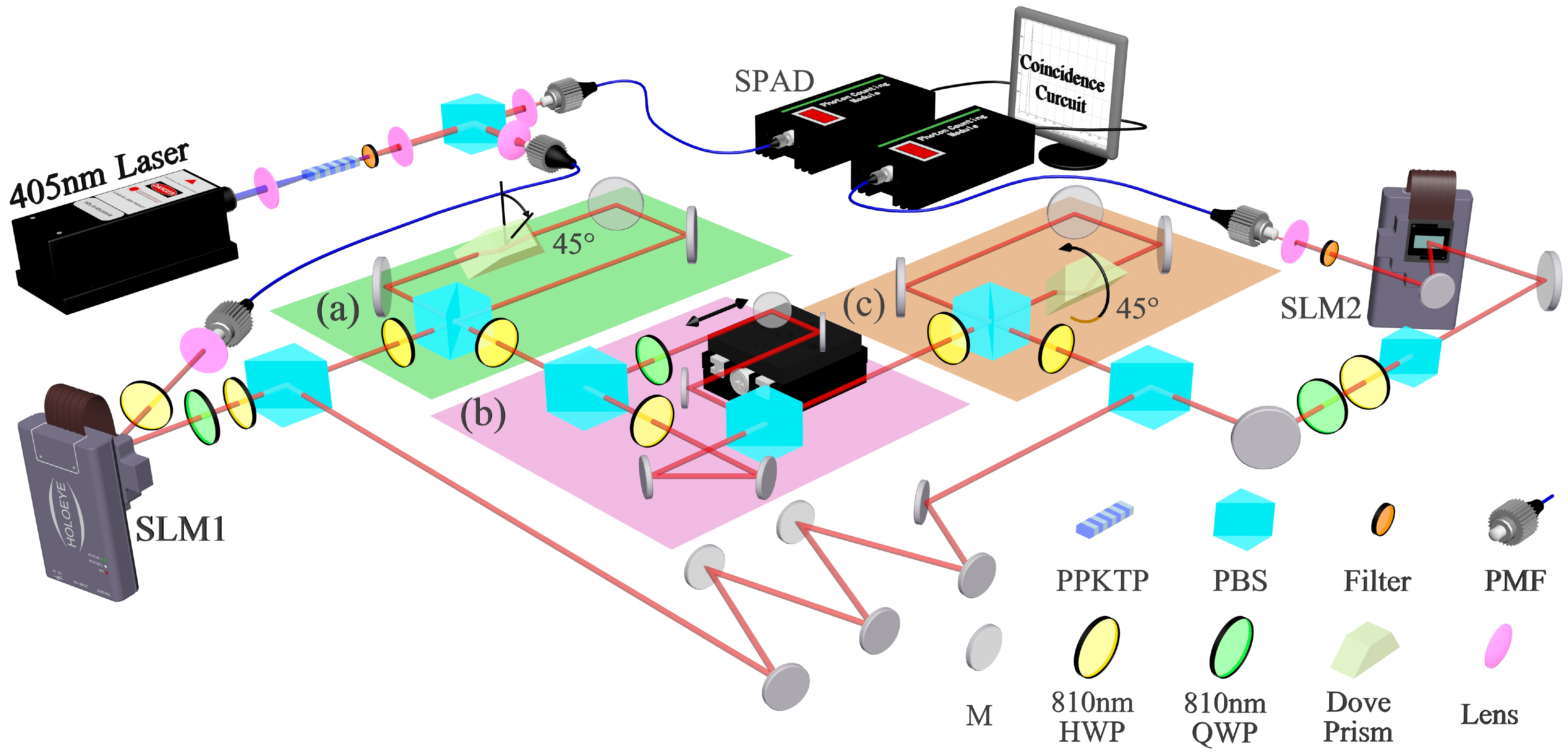}
\caption{Experimental implementation of the Toffoli gate. A Gaussian beam with a central 405 nm wavelength is focus on a 5 mm PPKTP nonlinear crystal to create 810 nm correlated photon pairs. Photons from the source are spatially filtered to the fundamental Gaussian mode by a PMF (blue wire). A HWP and the SLM1 can be used to separately modulate polarization and transverse modes of photons. (a)--(c) combined with two PBSs make up the Toffoli gate.
PPKTP: periodically poled potassium titanyl phosphate; HWP: half-wave plate; QWP: quarter-wave plate; PBS: polarizing beam splitter; SLM: spatial light modulator; PMF: polarization maintaining single-mode fiber; SPAD: single-photon avalanche detector.}\label{exp}
\end{figure}

Our experimental setup is shown in Fig.~\ref{exp}. Correlated photon pairs are created through a type-\Rmnum{2} spontaneous parametric down-conversion in 5-mm-long periodically poled potassium titanyl phosphate (PPKTP) crystal pumped by a 405 nm continuous-wave diode laser.
The final coincidence count for correlated photon pairs is 58.5 kHz with 12 mW average power, the coincidence efficiency is about 11.6\%.
One photon in a pair of the correlated photons is directly coupled into the polarization maintaining single-mode fiber (PMF) and detected by single-photon avalanche detector (SPAD) as a trigger signal.
The other is prepared for different quantum states to implement a Toffoli gate for an arbitrary input state.

Consider a three-qubit input state as follows:
\begin{align}
\left|\psi_0\right>&=(\cos\alpha_1\left|0\right>+e^{\mi\gamma_1}\sin\beta_1\left|1\right>)\notag\\
&\otimes(\cos\alpha_2\left|0\right>+e^{\mi\gamma_2}\sin\beta_2\left|1\right>)\notag\\
&\otimes(\cos\alpha_3\left|0\right>+e^{\mi\gamma_3}\sin\beta_3\left|1\right>).
\end{align}
It can be generated by the spatial light modulator (SLM1), a half-wave plate (HWP) and a quarter-wave plate (QWP), in which the coding method has been defined in Fig.~\ref{circuit}.
The polarizing beam splitters (PBSs) before and after Figs. \ref{exp}(a)--\ref{exp}(c) are used as the controlled operation for qubit 1.
Notice that for the down path (vertically polarized), there are eight reflections and a piezoelectric transducer (PZT, though not illustrated in Fig.~\ref{exp}) mounted on the last mirror to ensure identical optical length and no additional phase for the two paths.
Hence we only need to consider Figs. \ref{exp}(a)--\ref{exp}(c) as a $I\otimes U_{\mathrm{{\scriptscriptstyle CNOT}}}$ operator.
The initial state in up path can be written as follows:
\begin{align}
\left|\psi_1\right>&=\left|1\right>\otimes(\cos\alpha_2\left|0\right>+e^{\mi\gamma_2}\sin\beta_2\left|1\right>)\\
&\otimes(\cos\alpha_3\left|0\right>+e^{\mi\gamma_3}\sin\beta_3\left|1\right>)\notag\\
&=\cos\alpha_2\cos\alpha_3\left|100\right>+e^{\mi\beta_3}\cos\alpha_2\sin\beta_3\left|101\right>\notag\\
&+e^{\mi\beta_2}\sin\beta_2\cos\alpha_3\left|110\right>+e^{\mi(\beta_2+\beta_3)}\sin\beta_2\sin\beta_3\left|111\right>.\notag
\end{align}
And it should become
\begin{align}
\left|\psi_2\right>&=I\otimes U_{\mathrm{{\scriptscriptstyle CNOT}}}\left|\psi_1\right>\\
&=\cos\alpha_2\cos\alpha_3\left|100\right>+e^{\mi\beta_3}\cos\alpha_2\sin\beta_3\left|101\right>\notag\\
&+e^{\mi\beta_2}\sin\beta_2\cos\alpha_3\left|111\right>+e^{\mi(\beta_2+\beta_3)}\sin\beta_2\sin\beta_3\left|110\right>\notag
\end{align}
after Figs. \ref{exp}(a)--\ref{exp}(c).
It means that these four relationships, $\left|{100}\right>\to\left|{100}\right>$, $\left|{101}\right>\to\left|{101}\right>$, $\left|{110}\right>\to\left|{111}\right>$ and $\left|{111}\right>\to\left|{110}\right>$, are required through Figs. \ref{exp}(a)--\ref{exp}(c).
Meanwhile, Figs. \ref{exp}(a)--\ref{exp}(c) constitute a {\scriptsize CNOT} gate for qubits 2 and 3 encoded in four-dimensional OAM Hilbert space.

\begin{figure*}[!t]
    \centering
    \includegraphics[width=0.7\linewidth]{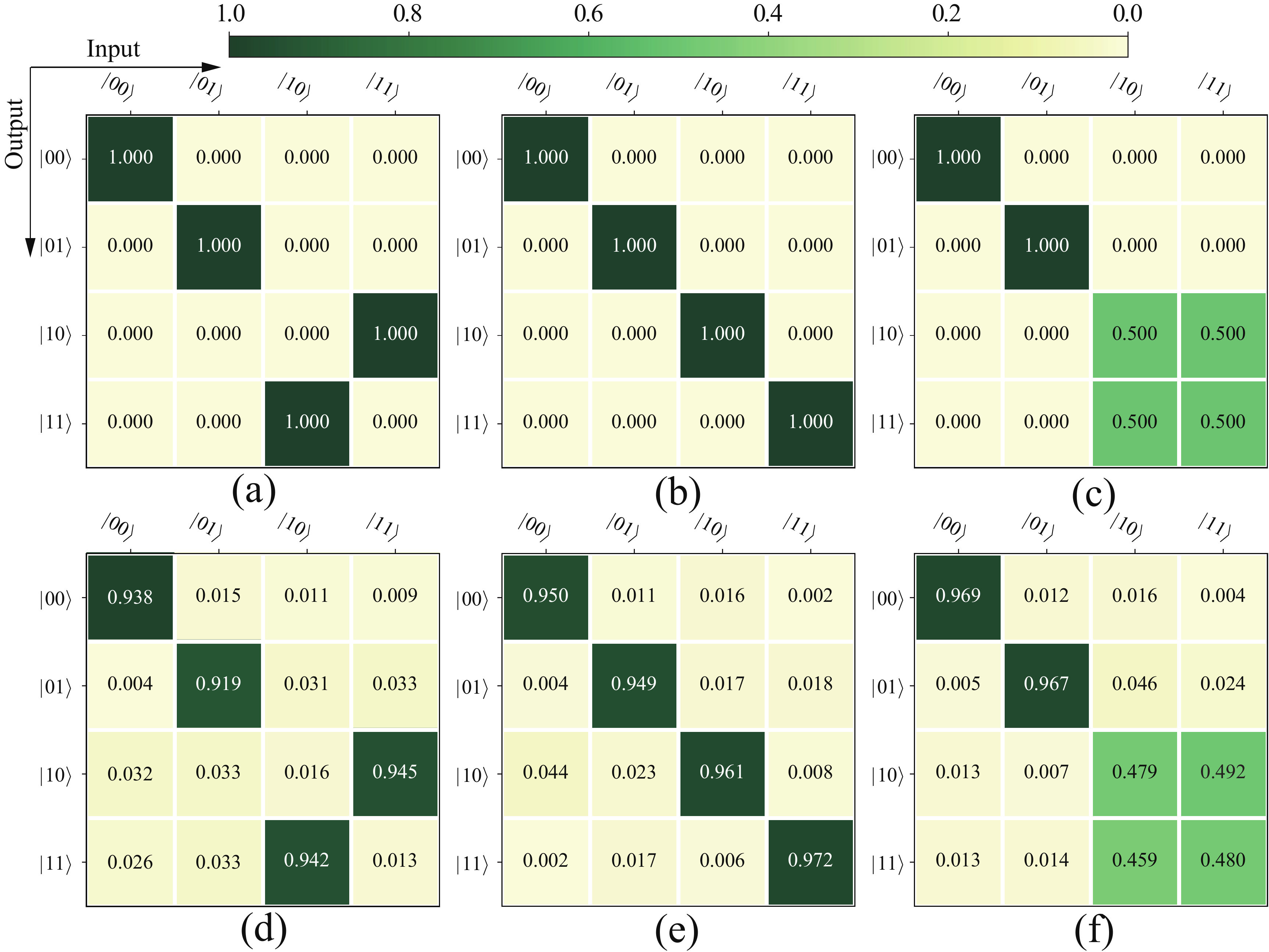}
    \caption{Experimental demonstration of Toffoli operation in the computational basis. (a) and (d) are the ideal computational basis operation and measured operation for the gate presented here while qubit 1 is at $\left|1\right>$ individually. (b) and (e) are the ideal and measured data while qubit 1 is at $\left|0\right>$ individually. (c) and (f) are the ideal and measured data while qubit 1 is at $(\left|0\right>+\left|1\right>)/\sqrt{2}$ individually.}\label{fig:cnot}
\end{figure*}

Therefore, in our experimental setup, the angles of the four HWPs sequentially in the Figs. \ref{exp}(a) and \ref{exp}(c) are $\pi/8$, $3\pi/8$, $\pi/8$ and $3\pi/8$, the angles of HWP and QWP in Fig. \ref{exp}(b) are $\pi/2$ and $0$, respectively,
and the angle of two Dove prisms are all $\pi/4$.
In addition, the Dove prism is to operate the OAM  state $\left|\ell\right>$ with the following rule,
\begin{align}
\left|\ell\right>\xrightarrow[]{\mathrm{Dove}} \mi e^{\mi2\alpha \ell}\left|-\ell\right>.
\end{align}
where $\alpha$ is the rotated angle of Dove prism.
Interestingly, Figs. \ref{exp}(a) and \ref{exp}(c) are two Sagnac interferometers, and are OAM parity sorters.
As for Fig. \ref{exp}(b), a motorized translation stage (MTS) and a PZT (on the reflector panel after MTS) are used to balance the optical path of two arms.
The target qubit is flipped by this Mach-Zehnder interferometer according to the different number of reflections in odd and even OAM modes.
Based on the above experimental settings, the four transformations are given as
\begin{subequations}\label{eq4}
\begin{align}
&\left|100\right>\xrightarrow[]{\mathrm{(a)}}\left|100\right>\xrightarrow[]{\mathrm{(b)}}\left|100\right>\xrightarrow[]{\mathrm{(c)}}\left|100\right>,\label{eq4a}\\
&\left|101\right>\xrightarrow[]{\mathrm{(a)}}-\left|101\right>\xrightarrow[]{\mathrm{(b)}}-\left|101\right>\xrightarrow[]{\mathrm{(c)}}\left|101\right>,\label{eq4b}\\
&\left|110\right>\xrightarrow[]{\mathrm{(a)}}\mi\left|010\right>\xrightarrow[]{\mathrm{(b)}}-\mi\left|011\right>\xrightarrow[]{\mathrm{(c)}}\left|111\right>,\label{eq4c}\\
&\left|111\right>\xrightarrow[]{\mathrm{(a)}}\mi\left|011\right>\xrightarrow[]{\mathrm{(b)}}-\mi\left|010\right>\xrightarrow[]{\mathrm{(c)}}\left|110\right>\label{eq4d}.
\end{align}
\end{subequations}
That is to satisfy $\left|\psi_2\right>=I\otimes U_{\mathrm{{\scriptscriptstyle CNOT}}}\left|\psi_1\right>$.

Since the SLMs in our experiment only respond to horizontally polarized photons, a series of wave plates and PBS before SLM2 convert incident photons into horizontally polarized photons. The SLM2 together with a PMF is used to perform any directional projection measurements of OAM modes.
After flattening its phase, an incoming photon is transformed into a Gaussian mode that can be efficiently coupled into the PMF.
Two SPADs record photon counts after the spin and OAM measurements, and their coincidence counts are proportional to the detecting correlated photon pairs.

% section Experimental_setup (end)

\section{Measurement Scheme and Experimental Results} % (fold)
\label{sec:measure}
To illustrate the quality of the Toffoli gate, we define the conversion rate $P_{ij}$ for the transformation.
It is obtained from the crosstalk measurements between the input and output modes in the computational basis,
according to $P(ij)=N_{ij}/\sum_{ij}N_{ij}$,
where $N_{ij}$ signifies the coincidence count of the input mode $i$ being transformed into the output mode $j$ in the crosstalk matrix.

To characterize the operation of this gate in more detail, we measured the outputs of the gate for each of the four possible computational basis input states ($\left|00\right>$, $\left|01\right>$, $\left|10\right>$ and $\left|11\right>$) while qubit 1 is at different states.
The first row of the Fig.~\ref{fig:cnot} represents the predicted results, and the second are the experimental results.
While the state of qubit 1 is $\left|1\right>$, the unitary transformation of qubits 2 and 3 should be a {\scriptsize CNOT} gate, illustrated in Fig.~\ref{fig:cnot}(a) and Fig.~\ref{fig:cnot}(d).
The results of this logic gate are consistent with the theoretical values and the effective conversion rate is $93.6\pm1.7\%$.
The Fig.~\ref{fig:cnot}(b) and Fig.~\ref{fig:cnot}(e) show that the states of qubits 2 and 3 indeed are invariant while the state of qubit 1 is $\left|0\right>$.
The remaining two subgraphs show that the transformation $(I_4+U_{\mathrm{{\scriptscriptstyle CNOT}}})/2$ works well while the state of qubit 1 is $(\left|0\right>+\left|1\right>)/\sqrt{2}$.

\begin{figure*}[!t]
    \centering
    \includegraphics[width=0.8\linewidth]{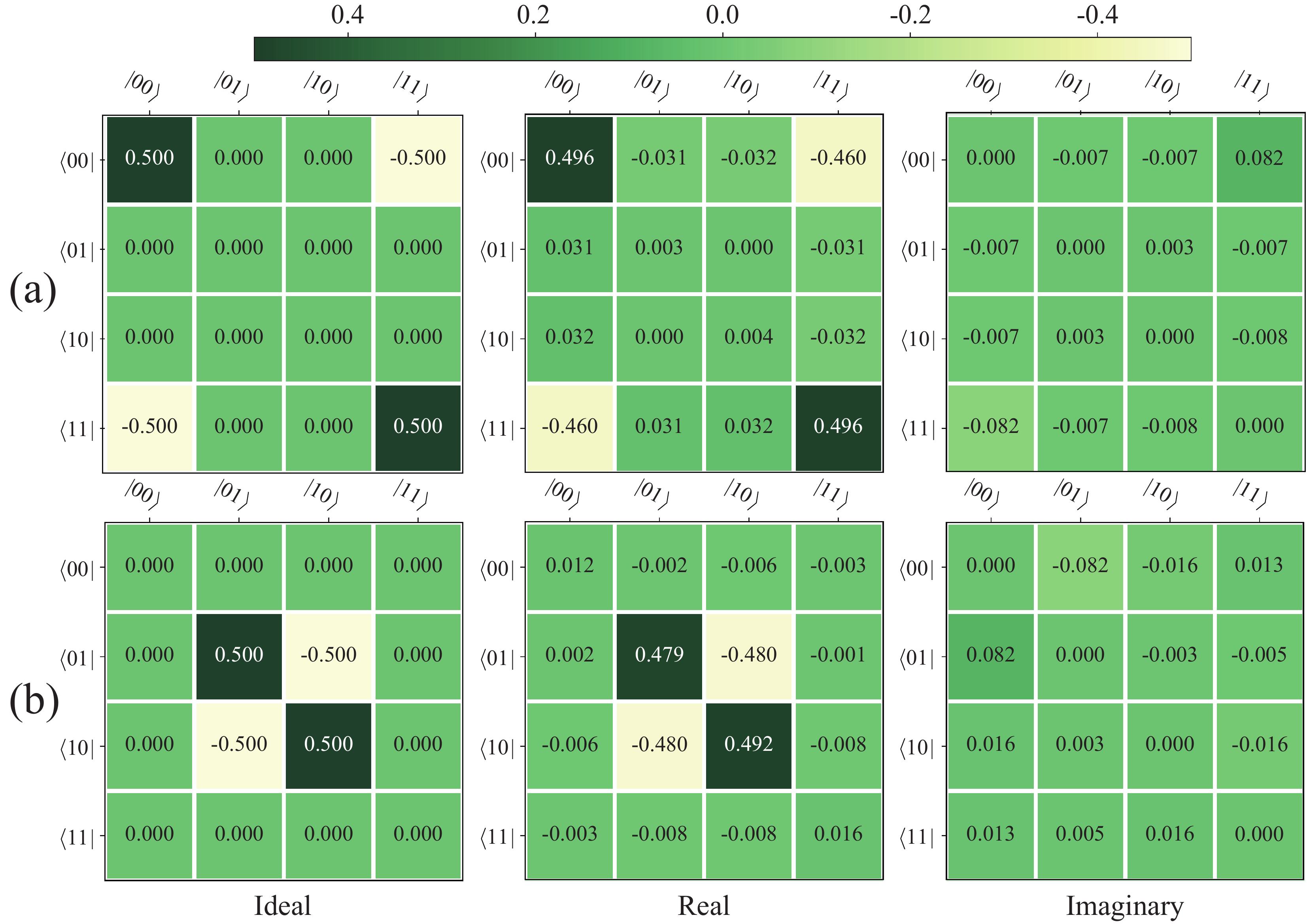}
    \caption{Density matrices for two  maximal entangled states. (a) The real part density matrix for the  maximal entangled Bell singlet state $\left|\Phi_-\right>$ (the imaginary components are all zero), and the real and imaginary parts of the density matrix reconstructed from quantum state tomography. (b)The real  part of the $\left|\Psi_-\right>$ state, and the real and imaginary parts of the density matrix reconstructed from quantum state tomography.}\label{fig:MES}
\end{figure*}

\begin{figure}[t]
    \centering
    \includegraphics[width=\linewidth]{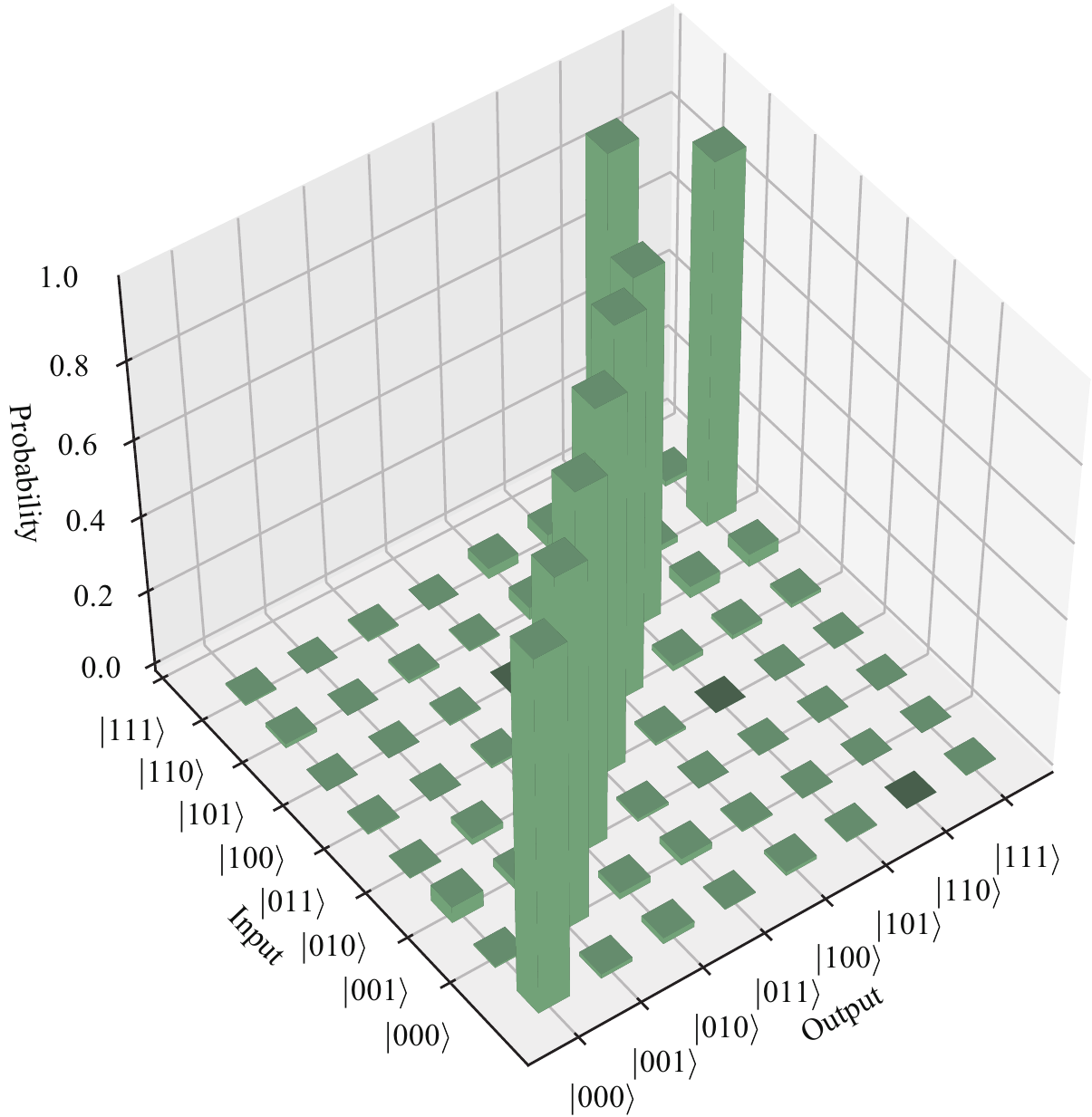}
    \caption{Truth table of the Toffoli gate. After preparing the three qubits in one of the eight basis input states $\left|000\right>$ to $\left|111\right>$, the probabilities of all basis output states are measured 100 seconds. The effective conversion rate of the truth table is $(95.1\pm3.2)\%$.}
    \label{fig:Tof}
\end{figure}

\begin{table}[!htbp]
\centering
\caption {Characterizations of {\scriptsize CNOT} gate with the four specific separable states.}
\label{tab:MES}
\begin{ruledtabular}
\begin{tabular}{ccc}
Input states & Theoretical output states &  Fidelity  \\
\hline
$(\left|0\right>+\left|1\right>)/\sqrt{2}\otimes\left|0\right>$&$\left|\Phi_+\right>=(\left|00\right>+\left|11\right>)/\sqrt{2}$&$96.9\pm2.3\%$\\
$(\left|0\right>-\left|1\right>)/\sqrt{2}\otimes\left|0\right>$&$\left|\Phi_-\right>=(\left|00\right>-\left|11\right>)/\sqrt{2}$&$95.6\pm1.6\%$\\
$(\left|0\right>+\left|1\right>)/\sqrt{2}\otimes\left|1\right>$&$\left|\Psi_+\right>=(\left|01\right>+\left|10\right>)/\sqrt{2}$&$96.5\pm1.8\%$\\
$(\left|0\right>-\left|1\right>)/\sqrt{2}\otimes\left|1\right>$&$\left|\Psi_-\right>=(\left|01\right>-\left|10\right>)/\sqrt{2}$&$97.1\pm2.1\%$\\
\end{tabular}
\end{ruledtabular}
\end{table}

\begin{table*}[!t]
\centering
\caption {The fidelity $F$ to detect a photon in input states $\rho_i$ though the Toffoli gate. The first column corresponds to the state of SAM qubit, and the first line corresponds to the state of OAM qubits.}
\label{tab:fidelity1}
\begin{ruledtabular}
\begin{tabular}{ccccc}
Input states & $\left|00\right>$ &  $\left|01\right>$ &  $\left|10\right>$&$\left|11\right>$  \\
\hline
$\left|1\right>$ & $93.8\pm2.7\%$ &  $91.9\pm2.8\%$ &  $94.2\pm2.1\%$&$94.5\pm1.9\%$\\
$\left|0\right>$ & $95.0\pm3.0\%$ &  $94.9\pm1.8\%$ &  $96.1\pm2.6\%$&$97.2\pm2.1\%$\\
$({\left|0\right>+\left|1\right>})/\sqrt{2}$ & $96.9\pm3.1\%$ &  $96.7\pm2.1\%$ &  $93.8\pm2.3\%$&$97.2\pm1.6\%$\\
\hline\hline
Input states & $\left|\Psi_+\right>$ &  $\left|\Psi_-\right>$ &  $\left|\Phi_+\right>$&$\left|\Phi_-\right>$  \\
\hline
$\left|1\right>$ & $96.6\pm2.1\%$ &  $94.7\pm2.4\%$ &  $94.0\pm1.7\%$&$95.4\pm1.6\%$\\
$\left|0\right>$ & $96.3\pm1.1\%$ &  $96.7\pm0.8\%$ &  $97.6\pm1.2\%$&$97.2\pm1.5\%$\\
$({\left|0\right>+\left|1\right>})/\sqrt{2}$ & $96.4\pm1.3\%$ &  $97.3\pm2.7\%$ &  $94.5\pm2.4\%$&$96.8\pm2.6\%$\\
\end{tabular}
\end{ruledtabular}
\end{table*}

It is known that {\scriptsize CNOT}  and Toffoli gates are very useful for preparing multi-party entanglement.
According to its role in quantum circuits, if we apply the {\scriptsize CNOT} gate to several separable states, these states would be transferred to maximal entangled states.
Experimentally, we test a few specific input states of qubits 2 and 3 listed in Table~\ref{tab:MES}.
We measure the fidelity between the experimental state $\rho_{o}$ and the expected state $\rho_{e}=U\rho_{i}U^{\dagger}$ ($\rho_{i}$ is the input state) of the unitary transformation as $F(\rho_o,\rho_e)=\left(\mathrm{Tr}\sqrt{\sqrt{\rho_o}\rho_e\sqrt{\rho_o}}\right)^2$.
The errors are calculated by using Monte Carlo simulations assuming Poisson counting statistics.
Figure~\ref{fig:MES}(a) shows the experimental results for the input state $(\left|0\right>-\left|1\right>)/\sqrt{2}\otimes\left|0\right>$,  where the ideal result is $\left|\Phi_-\right>$ in theory, the real and imaginary parts are the reconstructed density matrix of the output state in experiment. We obtain the fidelity  $F_{\Phi_-}=95.6\pm1.6\%$.
As depicted in Fig.~\ref{fig:MES}(b), the ideal state is $\left|\Psi_-\right>$ in theory while the input state is $(\left|0\right>-\left|1\right>)/\sqrt{2}\otimes\left|1\right>$, the real and imaginary components are the reconstructed result of density matrix in experiment. Its fidelity is $F_{\Psi_-}=97.1\pm2.1\%$. The fidelities for the other two Bell states are listed in Table~\ref{tab:MES}.

In addition, to supplement the above discussion, we also tested some other quantum states listed in Table~\ref{tab:fidelity1} when the states of controlled qubit 1 are $\left|0\right>$, $\left|1\right>$ and $(\left|0\right>+\left|1\right>)/\sqrt{2}$.
The truth table depicted in Fig.~\ref{fig:Tof} shows the probability of all computational basis states after applying the Toffoli gate to each of the computational basis states. It reveals the characteristic properties of the Toffoli gate, namely that a {\scriptsize NOT} operation is applied on the target qubit if the control qubits are in the state $\left|11\right>$.
All of the data listed here were taken over one day. During this time period, we only changed the angles of HWPs and QWP after and before two SLMs and grating images loaded on SLMs for measurement, and its performance was  repeatable.
This demonstrates two important points: the gate is highly stable, and it does not require realignment during the measurements for different input states.
In addition, corrections have been made for accidental coincidence counts, which will reduce some errors in gate operation.
% section Rresults (end)

\section{Conclusions} % (fold)
\label{sec:Conclusions}
In summary, we have presented the implementation of a deterministic Toffoli gate based on the OAM and polarization on a single-photon level.
The operation is characterized by different inputs in computational basis.
Moreover, Bell states between the different qubits realized by the controlled {\scriptsize NOT} gate is examined as well while the first controlled qubit value is $\left|1\right>$. The quantum operations performed in our work not only possess high stability and conversion efficiency, but they can also be easily extended to deal with multi-qubit input states in higher Hilbert space.
Our research simplifies the implementation of Toffoli gate, and has some effects on quantum information processing tasks based on hybrid DoFs.
% section conclusion (end)

\section*{Acknowledgments}
This work was in part supported by the National Nature Science Foundation of China (Grants No. 11534008, No. 11804271, and No. 91736104), Ministry of Science and Technology of China (2016YFA0301404) and China Postdoctoral Science Foundation via Project No. 2020M673366.

\nocite{*}
%\bibliography{bib.bib}

\end{document}